\documentclass{epl}

\usepackage{amssymb,epsfig}

\title{Point force manipulation 
and activated dynamics\\
 of polymers adsorbed on structured substrates}
\shorttitle{Point force manipulation of polymers}

\author{P. Kraikivski, R. Lipowsky, and J. Kierfeld}
\institute{Max-Planck-Institut f{\"ur} Kolloid- und
  Grenzfl{\"a}chenforschung, 14424 Potsdam, Germany}


\newcommand{\cH}{{\cal H}}

\begin{document}

\maketitle

\begin{abstract}
   We study  the activated motion of  adsorbed polymers
which are driven  over a structured substrate  by 
 a localized {\em point} force. 
  Our theory applies to  experiments with single polymers 
using, for example,  tips of scanning force microscopes 
 to drag the polymer.
We consider both flexible and semiflexible polymers, and 
the lateral surface structure  is represented by
double-well or periodic potentials.  
The dynamics is governed by kink-like excitations for which we  
calculate shapes, energies, and critical point forces.
Thermally activated  motion 
  proceeds by the nucleation of a kink-antikink pair at the point
where the force is applied and subsequent diffusive
  separation of kink and antikink. In the  stationary state of the 
driven polymer, the collective kink dynamics can be described by an
one-dimensional symmetric simple exclusion process. 
\end{abstract}

\section{Introduction}

The thermally activated escape over potential barriers
under the influence of an external  force  has been 
first solved by Kramers for a point particle \cite{Kramers}.  Since
then this process  has been  extensively studied not only 
for point particles \cite{Haenggi} but also for 
extended objects such as elastic strings.
Examples are provided by condensed matter systems:
  dislocation motion in crystals
\cite{Seeger,KP70}, 
motion of flux lines in type-II
superconductors \cite{vortex}, or charge-density waves \cite{CDW}.
An analogous problem is the activated motion of 
polymers over a potential barrier, which has been considered 
both for flexible  \cite{sebastian} and semiflexible
 polymers \cite{KLK04, KLK05}.

In all of these previous studies,  the thermally activated 
motion  is induced  by  spatially  {\em uniform} forces which
 are applied to the whole polymer or elastic line.
In contrast, in the present article, we will address
 the thermally activated 
motion of  polymers over potential barriers 
in the presence of 
a {\em point} force which acts only locally on the polymer.
We will consider both flexible and semiflexible polymers.  

Our theoretical study  is motivated by 
experimental  advances in  the manipulation  and
visualization of  single polymers using optical \cite{Ashkin}
 and magnetic \cite{Strick} tweezers, or scanning force microscopy
\cite{SeverinRabe}.
In ref.~\cite{SeverinRabe} it has been demonstrated that 
these techniques  allow to experimentally  apply 
localized point forces to a  polymer adsorbed on a substrate. 
Polymers that are strongly adsorbed onto crystalline substrates 
 such as graphite or mica  experience   a spatially 
modulated  adsorption  potential reflecting the underlying 
crystal  lattice structure and giving rise  to  preferred 
orientations of the adsorbed polymer.
For such systems,
 the  dynamics of the adsorbed polymer is governed by thermal activation 
over the potential barriers of the surface potential.

One example of  polymers adsorbed on  a  structured surface are  
self-assembling polymer chains consisting of 
  long-chain alkanes and alkylated small
molecules  on crystalline substrates such as the basal plane of
graphite \cite{RabeBuchholz}.
The alkyl chains orient along the substrate axes thereby providing an
effective periodic adsorption potential. 
Also  biopolymers such as  DNA or polyelectrolytes can be oriented 
on the basal plane of graphite by using long chain alkanes as an oriented 
template layer   \cite{kurth02,SeverinRabe}. 
It has been demonstrated experimentally that these polymers can be 
manipulated  individually on the structured surface
by applying  
point  forces using  the tip of a scanning force microscope
 \cite{SeverinRabe}.

Our main results are as follows. 
At low forces, the dynamics of the polymer 
is governed by thermal activation and 
 nucleation of localized kink-like excitations
 as shown in fig.\ \ref{shape}.
We calculate the critical point force below which the 
polymer moves by thermal activation over the  barriers of the
adsorption potential.
The steady state of this  activated motion determines 
the profile and velocity of the moving polymer and 
is governed by the (collective) driven motion of 
the kink excitations which can be described as 
a one-dimensional symmetric simple 
 exclusion process of these excitations.
Our  results for the critical point force, the velocity, and the 
profile of the moving polymer are  accessible in
manipulation experiments on adsorbed polymers and allow 
to extract  material parameters of the  polymer and the substrate
structure from such experiments. 
We will first present detailed calculations for stiff, 
semiflexible polymers.  Results for flexible polymers 
are discussed in the end.

\section{Model}
\label{sec_model}

\begin{figure}[t]
 \begin{center} 
  \epsfig{file=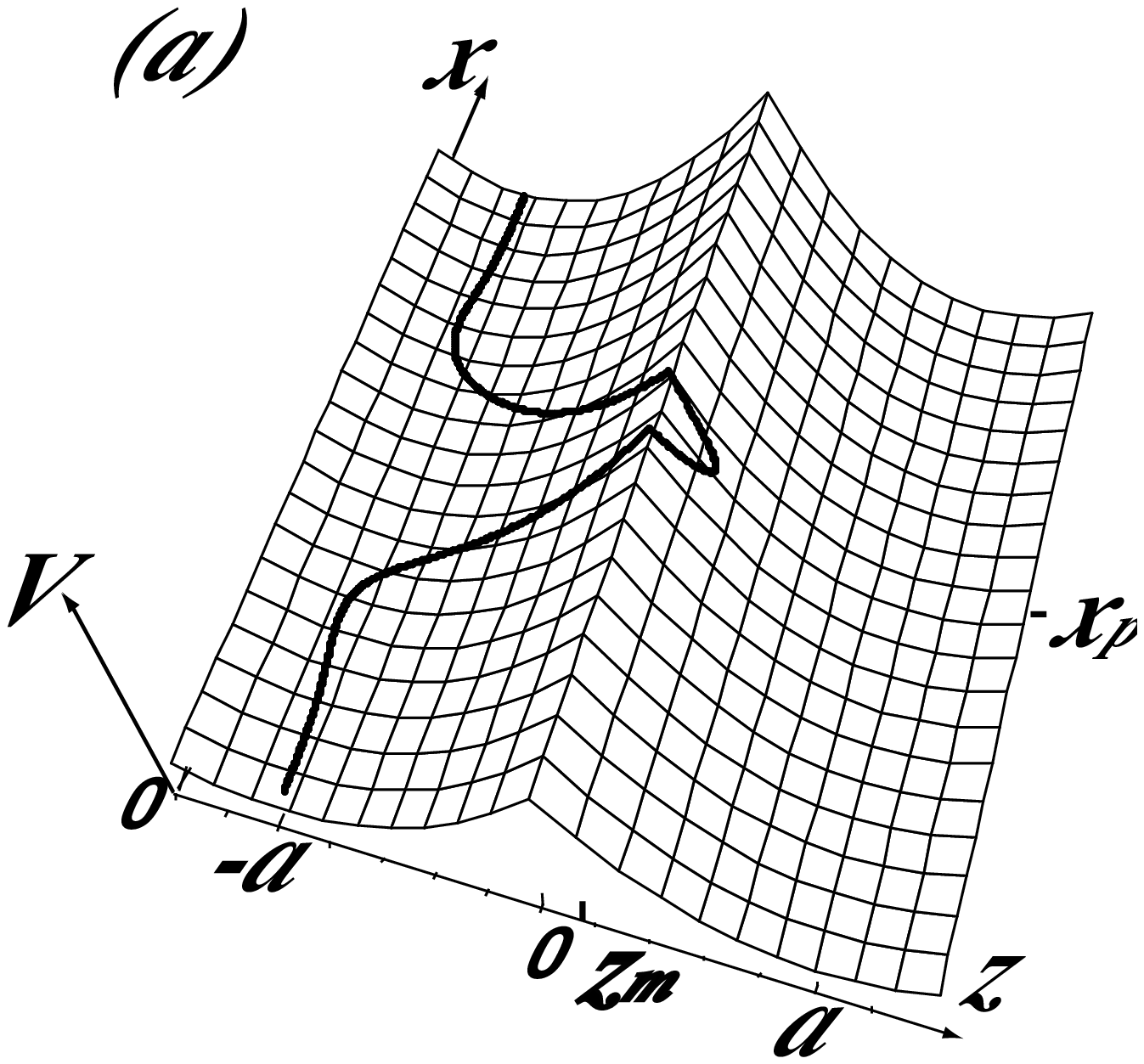,width=0.45\textwidth}~~~
 \epsfig{file=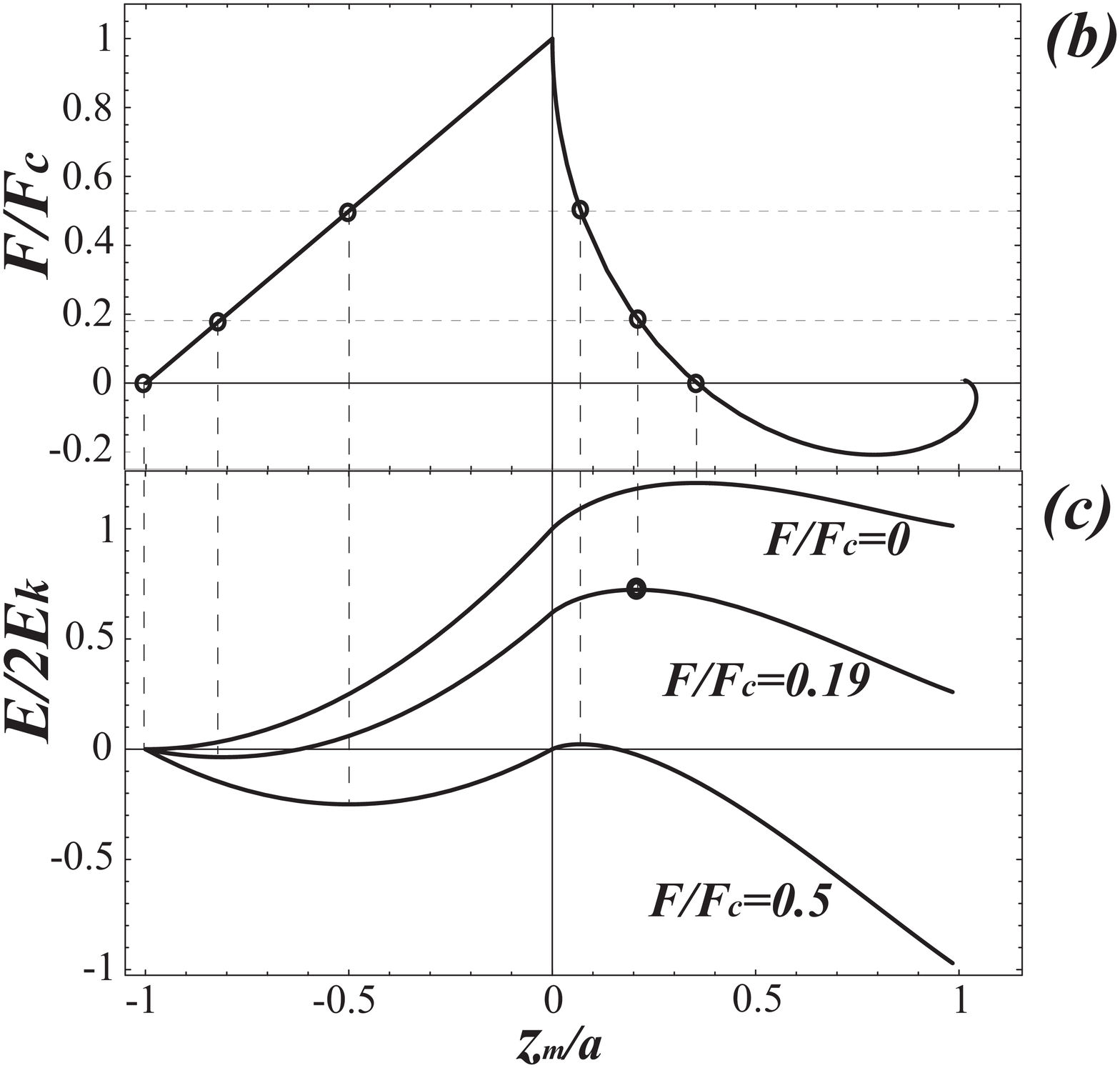,width=0.5\textwidth}
 \caption{    
    \label{shape}\label{zmEp}
(a) Kink-antikink
  configuration of a semiflexible 
 polymer in a double-well potential $V$  under
 the action of a point force $F_p$ displacing  the midpoint in
  the   $z$-direction to a value $z_m$. 
 The configuration $z_k(x)$ is calculated for 
  $F_p/F_{c}=0.19$, $z_m/a=0.21$
  ($L_2/w_k=1.1$, $L_1/w_k=15$) and has an energy $E/2E_k=0.72$.
(b)  The midpoint $z_m$ (in units of $a$) as a function of the external 
force (in units of critical force $4E_k/a$).
(c) Energy $E(z_m)$  (in units of $2E_k$) 
 of a kink-antikink pair  as a function of the 
 midpoint $z_m$ (in units of $a$) for different forces $F_p/F_c=0,0.19,0.5$.
 }
 \end{center}
 \end{figure}

We consider the dynamics of a semiflexible polymer adsorbed to a 
planar two-dimensional structured substrate  under  the influence of  an 
external point force $F_p$ pulling the polymer. A generic model of 
the substrate structure is  a double-well potential that is 
translationally invariant  in 
one direction, say the $x$-axis as in fig.\ \ref{shape}.
The semiflexible polymer has a bending rigidity $\kappa$ and
persistence length $L_p = 2\kappa/T$ where $T$ is the temperature in
energy units. We focus on the regime where the  potential 
wells are sufficiently deep and narrow
so that the adsorbed polymer is oriented along the  $x$-axis  and   can be 
parameterized by displacements $z(x)$ perpendicular to the $x$-axis
with $-L/2 < x < L/2$, where $L$ is the projected length of
polymer, see fig.\ \ref{shape}.
The Hamiltonian of an oriented  polymer is given by
\begin{equation}
\cH\{z(x)\} = \int_{-L/2}^{L/2} dx
  \left[  \frac{\kappa}{2}\left(\partial_x^2z\right)^2+ V(z)
    \right] 
~, 
\label{hamil}
\end{equation}
i.e., the sum  of  bending 
and   potential energy \cite{KLK04, KLK05}. 
We consider a piecewise harmonic double-well potential
\begin{equation}
 V_p(x,z)\equiv  V_0(z) -F_p\delta(x-x_p)z 
\label{Vp}
\end{equation}
with $V_0(z) \equiv  \frac{1}{2} V_{0}(|z|-a)^2$, where  
$V_0$ is the depth of the potential. 
The potential (\ref{Vp})
 contains the action of a point force pulling the
polymer at the point $x=x_p$ with a force $F_p$ 
 in the $z$-direction. 
For zero point force $F_p=0$,  
the potential is symmetric, translationally invariant in
the $x$-direction, has  a 
 barrier height $V_0a^2/2$,  and the  distance 
between minima is $2a$. For $F_p>0$, the point force 
in (\ref{Vp}) breaks the translational invariance 
of the system.

Our assumption of  an oriented polymer is valid if U-turns of the
polymer  
within a single potential well are suppressed by the bending energy. 
This is the case if the size $2a$ 
of each potential well in the $z$-direction 
 is  smaller than the  
 persistence length $L_p$. This condition is typically fulfilled 
for adsorbing substrates structured  on the nm-scale \cite{RabeBuchholz}.
Furthermore, the 
polymer should be  strongly adsorbed, which
corresponds to  a small 
density of thermally induced kink excitations, i.e., 
 $E_k\gg T$  where $E_k$ is the 
kink energy, see eq.~(\ref{Ek}) below and ref.\ \cite{KLK04} .

The overdamped motion of  the polymer is described by \cite{KLK04, KLK05}
\begin{equation}
   \gamma \partial_t z
  =  -\frac{\delta \cH}{\delta z} + \zeta(x,t)
  = - \kappa\partial_x^{4} z
    -V_0'(z) +F_p\delta(x-x_p) +  \zeta(x,t)
~,
\label{EOMf}
\end{equation}
 where $\gamma$ is the damping constant and $\zeta(x,t)$ is a 
Gaussian distributed thermal random force
with $\langle \zeta \rangle=0$ and
correlations $\langle \zeta(x,t) \zeta(x',t') \rangle 
=2 \gamma   T\delta(x-x')\delta(t-t')$. 
We neglect longitudinal motion of polymer segments
(see ref.\cite{KLK05} for a discussion)  and do not study 
the effects of an external tension or compression. 
For $V_p=0$,  tension and  compression have been considered in \cite{S96}
and \cite{G98}, respectively.

\section{Static kinks}
\label{static_kink}

First, we calculate the stationary shape of the semiflexible 
polymer that is
deformed by a point force acting 
at its midpoint  into a 
kink-antikink 
configuration  $z_{k}(x)$ as shown in fig.~\ref{shape}a.
This  configuration is obtained by displacing the 
polymer at the midpoint where the point force acts to a 
prescribed position  $z_m$  and letting the 
rest of the polymer equilibrate. 
Therefore, we have to solve
 the saddle-point equation ${\delta \cH}/{\delta z}=0$ for the
energy (\ref{hamil}), i.e.,  eq.\
(\ref{EOMf}) for the  time-independent case and in the absence of
noise ($\zeta=0$),  with appropriate  boundary conditions {\em and}
a  prescribed position $z_k(x_p)=z_m$.
For $z_m>0$ the kink configuration crosses the barrier at two points, 
see fig.~\ref{shape}a;
we choose  the origin $x=0$ and the  length $L_2$ 
such that these points are $z_{k}(0)=0$  and $z_{k}(L_2)=0$. 
The polymer  has a total length  $L=L_1+L_2$
and  extends from $x=-L_1/2$ to $x=L_1/2+L_2$, and the force 
acts at the midpoint $x_p = L_2/2$.
The kink-like configuration has to fulfill four
boundary  conditions,  $z_{k}(-L_1/2)=z_{k}(-L_1/2+L_2)=-a$
 and 
 $z'_{k}|_{-L_1/2}=z'_{k}|_{-L_1/2+L_2} =0$.
At the midpoint $x_p=L_2/2$,  we fix 
the displacement $z_m$ of the polymer $z_{k}(x_p)=z_m$,  and 
the  point force   causes a discontinuity
 in the  third derivative, 
$z'''_{k}(x_p+)-z'''_{k}(x_p-)=F_p/\kappa$. 
In addition, $z_{k}(x)$ and its first two derivatives have to be
continuous at the midpoint,  and  $z_{k}(x)$ and its first 
three derivatives have to be continuous at each crossing point
 $x_0 = 0,L_2$.

Away from the point force, i.e., for $x\neq x_p$
 the saddle point solutions 
are linear combinations 
of the  four  functions 
$\exp({\pm}x/w_{k})\exp({\pm}ix/w_{k})$ where $w_k\equiv
 \sqrt{2}(\kappa/V_0)^{1/4}$ is the kink width. 
Construction of the solution through the four regions  separated by
 the crossing points and the midpoint then 
 requires  to determine
 16 linear expansion coefficients and the two  parameters 
$L_2$ and $z_m$ as a  function of 
the system size $L$ and
the remaining model parameters including the point force  from 
the boundary and matching conditions. 
The resulting shapes of the  kink-like polymer configurations 
are shown in fig.~\ref{shape}a. 
fig.~\ref{zmEp}c  shows the energies $E(z_m)$ of the kink-like 
configuration as a function of $z_m$ for different point forces $F_p$. 
For low forces the energies $E(z_m)$ in fig.~\ref{zmEp}c
have  two stationary  points, a 
stable minimum at  $z_m=z_{m,min}<0$ 
 (the midpoint does not cross the barrier) 
and an unstable maximum at $z_m=z_{m,nuc}>0$.
This maximum is unstable with respect to further displacement of the
 midpoint and represents the critical nucleus configuration. 
For $F_p=0$, we obtain another stable minimum at $z_m=a$ 
(the midpoint reaches the next potential well)
which is the  static kink-antikink solution \cite{KLK04}.
 The width $w_k$
of a static  kink and its characteristic energy $E_k$   are given by 
\begin{equation}
  w_k= \sqrt{2}(\kappa/V_0)^{1/4}
~~,~~~
 E_k=a^2\kappa^{1/4}V_0^{3/4}/\sqrt{2}
~.
\label{Ek}
\end{equation}
In the limit of large $L$, we can find analytic expressions for the 
resulting stationary positions $z_{m,min}$ and  $z_{m,nuc}$ 
 as a function of the applied 
force $F_p$, see   fig.~\ref{zmEp}b.
We find  that there are no stationary positions  
if the  point force $F_p$ exceeds a critical value
$F_{c}$ given by 
\begin{equation}
F_c = 4E_k/a =  
    2\sqrt{2}a\kappa^{1/4}V_0^{3/4}
.
\label{Fc}
\end{equation}
The midpoint displacement  $z_{m,min}<0$ in the 
stationary minimum  is  a 
linear function of the external force, $z_{m,min} = -a\left(1-
F_p/F_c\right)$ and reaches the barrier at $z_{m,min}=0$ for
$F_p=F_c$,
see   fig.~\ref{zmEp}b.
This  force-displacement relation
 describes  the linear response of the 
polymer before crossing the barrier. 
For the midpoint displacement in the unstable nucleus 
configuration $z_{m,nuc}>0$, on the other hand, 
we obtain the  following set of two equations for 
$z_{m,nuc}$ and $L_2$,
\begin{equation}
{F_p}/{F_c} = 
\left. (\cos x- \sin x)e^{-x}\right|_{x=L_2/2w_k} 
~~,~~
z_{m,nuc}/a = \left. 1- (\sin x+\cos x)e^{-x} \right|_{x=L_2/2w_k} 
~.
\label{zmL2}
\end{equation}
As shown  in  fig.~\ref{zmEp}b, 
$F_p$ is decreasing for increasing $z_{m,nuc}$ as 
the critical nucleus configuration widens for small point forces. 
The negative values of $F_p$ for large $z_{m,nuc}$ indicate that 
for a semiflexible polymer 
the kink-antikink configuration reached for $z_m=a$ is stabilized
 by an energy barrier. Only below a 
negative threshold  force  $F_c^-\equiv -F_c e^{-\pi/2} <0$  the 
 kink-antikink configuration  becomes unstable.

\section{Kink nucleation}
\label{nucl}

Now we turn to the
activated kink  nucleation in the presence of 
 a point force pushing the polymer over the potential barrier. 
The point force breaks the translational invariance in $x$-direction and 
kink-antikink pairs are only nucleated at $x=x_p$ with a rate 
$J$ per unit time. 
This thermally activated 
process is governed
by an energy barrier which is given by the excess energy $\Delta
E_{n}$ of the critical nucleus configuration. 
The energy of the critical nucleus can be obtained  from 
the  energy profiles  $E(z_m)$ shown in 
 fig.~\ref{zmEp}c as 
 the difference  $\Delta E_n\equiv E(z_{m,nuc}) -
E(z_{m,min})$ between minimum and maximum values of the 
energy  $E(z_m)$  of the kink-like configuration. 
We find 
$\Delta E_{n}\sim 2E_k\left(1-{F_p}/{F_{c}}\right)^2$,
which vanishes
as the force approaches the  critical value
$F_{c}$.
The activation energy 
 enters the nucleation current 
\begin{equation}
 J = (Q_{n}/2\pi)\exp\left(-\Delta E_n/T\right)
~~~~\mbox{with}~~~
Q_{n}^2 \equiv |\omega_{n,0}|\omega_{s,0}
   \prod_{p>0}(\omega_{s,p}/\omega_{n,p}) 
 ~,
\label{currentFp}
\end{equation}
 which shows Arrhenius-type
behaviour. The  prefactor $Q_{n}$ 
includes the spectrum of 
attempt frequencies $\omega_{n,p}$ and $\omega_{s,p}$ ($p=0,1,...$)
for phononic fluctuations around the critical nucleus configuration 
and the straight  configuration $z_m=-a$, respectively. 
We find one unstable negative mode $\omega_{n,0}\leq 0$, 
which diverges as  $\omega_{n,0}=(V_0/\gamma)
   [1-2^{4/3}(1-F_p/F_{c,\kappa})^{-4/3}]$ upon approaching the
critical force $F_p\approx F_c$, 
a bound state with $0<\omega_{n,1}\leq V_0/\gamma$, 
and a set of positive modes  $\omega_{n,p}>V_0/\gamma$
 with the same level spacing  as  the modes of the straight configuration.
It is important to note that 
two   translational modes (for kink and antikink) 
  only exist if  the point force is zero
 because the  point force breaks the translation invariance. 
Close to the critical force $F_p\simeq  F_c$, 
we obtain  $Q_n^2\approx
(V_0/\gamma)^2[1-2^{4/3}(1-F_p/F_c)^{-4/3}]$.

\begin{figure}[t]
 \begin{center}
   \epsfig{file=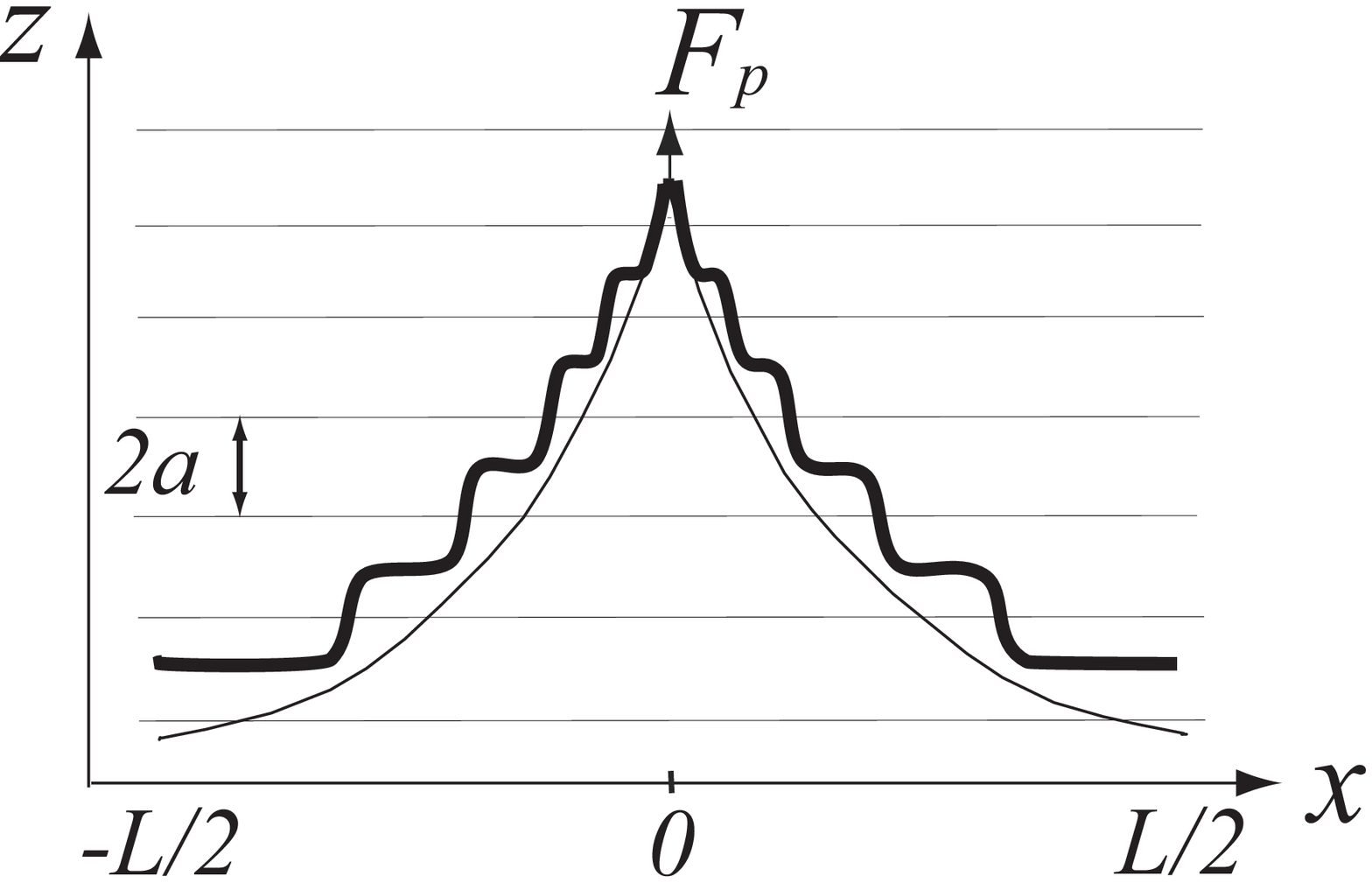,width=0.38\textwidth}~~~~~
 \epsfig{file=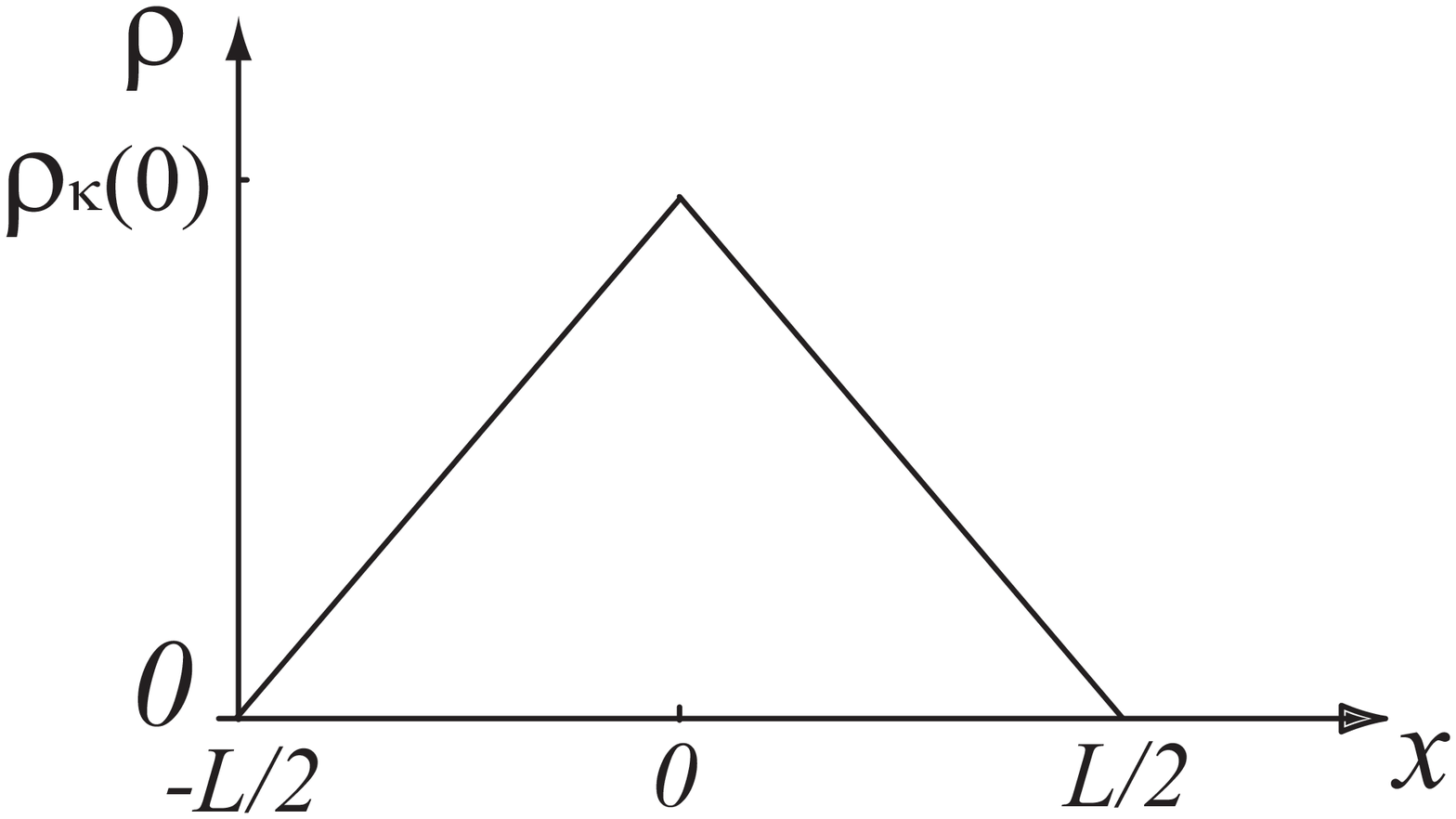,width=0.38\textwidth}
 \caption{\label{dprofile}
(Left) The shape of a semiflexible  polymer pulled 
over a periodically structured surface 
by a  point force acting at the 
midpoint. The horizontal lines indicate the position of
potential barriers. The thick solid line shows a typical polymer 
configuration $z(x)$, the thin line the average shape $\langle z(x)
   \rangle$. 
(Right) The stationary  kink density $\rho_k(x)$ 
as a function of the distance from the
   point $x_p$ where 
the force is acting on the polymer. 
 }
 \end{center}
 \end{figure}

\section{Collective kink dynamics}

After nucleation of a kink-antikink pair 
  at $x=x_p$ by thermal activation,
 kink and antikink are driven apart by a small  force 
$\sim E_k/w_k e^{-L_2/w_k}$,
 which  decays exponentially with the 
distance $L_2>w_k$ between kink and antikink. 
This exponential decay is characteristic for a  
 point driving force which  interacts only 
over a distance $\sim w_k$ with the kink and  very different 
from the case of  a spatially  uniform   force, where 
 kinks experience a spatially uniform driving 
force \cite{BL79,KLK04}. 
For separations $L_2 > w_k$   the kink diffuses essentially  freely 
with a diffusion constant 
$D_k=2Tw_k/3\gamma a^2$ \cite{KLK04}.

A spatially localized 
 driving force also leads to a distinct steady state motion 
of the polymer in a periodically continued potential, see  fig.\ 
\ref{dprofile}. This motion can be described in terms of 
 the collective dynamics of an ensemble of kinks and antikinks 
which are  generated at the  {\em single} point 
$x=x_p$ by the point force and subsequently separated by the
exponentially  decaying force. For the following discussion we 
choose coordinates such that 
 $x_p=0$, and the polymer extends from $-L/2<x<L/2$.
Because a point force creates kink-antikink pairs only at $x=0$, 
we find an  ensemble 
consisting only of kinks  in the region $x>0$ and an ensemble
consisting only of antikinks in $x<0$.  
As two (anti-)kinks  have a mutual short-range 
repulsion of range $w_k$, we have an 
ensemble of diffusing kinks (antikinks) with a hard-core  repulsion 
on the interval $L/2>x>0$ ($-L/2<x<0$).
In order to treat the  non-equilibrium dynamics of these ensembles,  
we introduce a discrete  one-dimensional lattice of possible kink 
positions with spacing $\Delta x = w_k$ which allows to 
 map the dynamics of each  ensemble onto 
the  {\em  symmetric simple exclusion process}  (SSEP)
with open boundaries \cite{spohn,schuetz}. 
In the following we consider the kink ensemble ($x>0$); the antikink ensemble 
($x<0$) can be treated analogously. 
In the kink ensemble, the kink particles are freely diffusing, i.e., 
they have {\em symmetric} rates $D\equiv D_k/w_k^2$ for hopping to the right 
and left on the lattice $x_i= iw_k$ ($i=1,...,N$ with $N=L/2w_k$); 
they  interact through their hard-core repulsion.  
In the SSEP, boundary conditions are specified by rates 
$\alpha$ and $\delta$ for particles  to enter the system at 
the left ($i=1$) and right ($i=N$), respectively, if that site is
empty.
For the kink ensemble we have $\alpha=J$,  as kinks are nucleated  
at  $i=1$ with the Kramers rate (\ref{currentFp}), and $\delta=0$ as 
no kinks enter the system at  $i=N$.
Furthermore, 
kinks leave the system diffusively,  at $i=1$ by
annihilation with an antikink and at $i=N$ by  
relaxation of the free  polymer end.

Despite the hard-core interaction the stationary density profile 
 $\rho_k(x)$ of kinks in the SSEP fulfills the stationary diffusion equation, 
$\partial_x^2 \rho_k =0$ \cite{spohn,schuetz}. 
Furthermore, our boundary conditions  are equivalent to boundary conditions 
$\rho_k(0) = w_k^{-1}\min(\alpha/D,1)$ and $\rho_k(L/2)=0$ for the 
stationary kink density at the  ends of the system. 
For $\alpha>D$ the system reaches its maximal kink density $w_k^{-1}$ 
at $x=0$. 
The resulting  linear  density profile $\rho_k(x)$ is
\begin{equation}
 \rho_k(x) =\rho_k(0) (1-2|x|/L)
  ~~~\mbox{with}~~~
 \rho_k(0) = w_k^{-1}\min(\alpha/D,1) = \min(Jw_k/D_k,1/w_k)
\label{rhokx}
\end{equation} 
as shown in  fig.\ \ref{dprofile} (right). 
The average distance between kinks is $1/\rho_k(x)$ and 
 at each kink the polymer position changes by  
$\Delta z=-2a$ leading to a characteristic 
{\em parabolic}   polymer shape 
$  \langle z(x) \rangle -z_m  =   - 2a\int_0^{|x|}
      d\tilde{x}\rho_k(\tilde{x}) 
     = - 2(a/w_k) \min(Jw_k^2/D_k,1) |x|(1-|x|/L)$ in the stationary state 
as shown in fig.\ \ref{dprofile} (left). 
The average velocity $v_z\equiv \langle \partial_t z \rangle$ 
of  the polymer in the $z$-direction is determined by the 
stationary current 
$J_{\rm SSEP}=-D_k \partial_x\rho_k =\min(J,D_k/w_k^2) w_k/L$
of  the SSEP. Only for 
 small nucleation rates $J \ll D_k/w_k^2$ the kink interaction can be
neglected and the current is directly given  by the Kramers 
rate (\ref{currentFp}), $J_{\rm SSEP} \approx Jw_k/L$.
 During the  time $1/J_{\rm SSEP}$ the polymer advances by 
a distance $2a$ leading to  $v_z =  2aJ_{\rm SSEP}\approx 
2a\min(J,D_k/w_k^2) w_k/L$.

\section{Flexible polymers}

So far we considered  semiflexible  polymers 
dominated by their bending energy. In this section we want to outline
the main results for  flexible 
Gaussian polymers governed by entropic elasticity 
with a tension  $\sigma = 2T/b$ 
where $b$ is the Kuhn length.
 The Hamiltonian of a 
flexible Gaussian polymer on a planar  two-dimensional 
substrate  is given by
\begin{equation}
\cH = \int_{-L_c/2}^{L_c/2} ds
  \left[\frac{\sigma}{2}\left[(\partial_s x)^2+(\partial_s z)^2\right]
+ V(z)
    \right] 
~, 
\label{hamilf}
\end{equation}
where we integrate over the arc length $s$ 
with $-L_c/2<s<L_c$, and  $L_c$ is the contour length of the polymer.
The translationally invariant 
potential $V(z)$ is a function of $z$ only.  Therefore
fluctuations in the $x$-coordinate decouple and  are  Gaussian with 
moments   $\langle (x(L_c)-x(0)) \rangle=0$
and $\langle (x(L_c)-x(0))^2 \rangle \approx L_cb/2$. 
The Rouse  dynamics of the $z$-coordinate of the polymer 
is given by 
\begin{equation}
   \gamma \partial_t z
=\sigma\partial_s^{2} z -V_0'(z) +F_p\delta(s-s_p) +  \zeta(s,t)
~,
\label{EOMff}
\end{equation}
where $\gamma$ is the damping constant and $\zeta(x,t)$ is a 
Gaussian distributed thermal random force. 
The point force is on the monomer $s=s_p$. 
For a flexible polymer the kink width is  $w_{k,\sigma}=(\sigma/V_0)^{1/2}$ and
the kink energy $E_{k,\sigma}=a^2(\sigma V_0)^{1/2}$ \cite{BL79,sebastian}. 
As for the semiflexible polymer we can calculate the energy $E(z_m)$ of a
kink-antikink configuration with prescribed midpoint $z_k(s_p)=z_m$. 
For a flexible polymer the displacements in the 
stationary minimum at $z_m=z_{m,min}$ and the
maximum representing the nucleus with  $z_m=z_{m,nuc}$ are both 
{\em linear} functions of the external force, 
$z_{m,min} = - a\left(1- F_p/F_{c,\sigma}\right)$ and 
$z_{m,nuc} =  a\left(1- F_p/F_{c,\sigma}\right)$, where the critical force 
for the flexible polymer is given by $F_{c,\sigma}=2E_{k,\sigma}/a
=2a(\sigma V_0)^{1/2}$. 
The nucleation
current $J_\sigma$ for the flexible polymer is given by
 the same expression (\ref{currentFp}) as for a semiflexible
polymer with the excess energy 
$\Delta E_{n}\sim 2E_{k,\sigma}\left(1-{F_p}/{F_{c,\sigma}}\right)^2$. 
The spectrum of attempt frequencies $\omega_{n,p}$ for the critical
nucleus shows slightly different
behaviour for the flexible polymer as the unstable negative mode 
 $\omega_{n,0} \approx -3V_0/\gamma$ does not diverge for  $F_p\approx
F_{c,\sigma}$, 
and we  finally obtain $Q_n\approx \sqrt{3}V_0/\gamma$.
The collective kink dynamics for a flexible polymer can also be mapped onto 
a one-dimensional SSEP. 
As a function of the  arc length $s$, 
we find a linear stationary kink density profile $\rho_k(s)=\rho_k(0)
(1-2|s|/L_c)$ with $\rho_k(0) =
\min(J_{\sigma}w_{k,\sigma}/D_{k,\sigma},1/w_{k,\sigma})$ 
and a parabolic
shape $\langle z(s) \rangle= - 2(a/w_{k,\sigma})
\min(J_{\sigma}w_{k,\sigma}^2/D_{k,\sigma},1)
|s|(1-|s|/L_c)$ 
 analogously to  the semiflexible polymer, cf.\ eq.~(\ref{rhokx}),
where  $D_{k,\sigma}=Tw_{k,\sigma}/\gamma a^2$ \cite{BL79} is the 
kink diffusion constant of the flexible polymer. 
In the real space 
 coordinates of the substrate, however,  the
resulting shape is $(\langle x(s) \rangle, \langle z(s) \rangle)= 
(0,\langle z(s) \rangle)$ and thus,  
 the parabolic shape is lost due to the decoupled Gaussian
fluctuations in the $x$-direction. 
The result  for the  velocity   $v_z =  2aJ_{\rm SSEP}\approx 
2a\min(J_{\sigma},D_{k,\sigma}/w_{k,\sigma}^2) w_{k,\sigma}/L_c$ 
is  analogous to the  semiflexible polymer.

\section{Conclusion}

In summary, we  described 
the activated motion of  single adsorbed polymers
on  a structured substrate displaced by localized {\em point} 
forces, which can be realized experimentally using, e.g., 
scanning force  microscopy tips. 
The dynamics is governed by kink-like excitations for which we have 
calculated shapes, energies, and critical point forces. 
Kink and antikink pairs  
are {\em locally} nucleated by the point force and then undergo 
a separation which is diffusive on separations larger than 
the kink width $w_k$. 
We have calculated the nucleation rate (\ref{currentFp}) 
using Kramers theory. 
The collective kink dynamics can be mapped onto 
a one-dimensional symmetric simple 
exclusion process (SSEP). Using this mapping we find
the average polymer velocity and a characteristic  
average parabolic shape for a driven semiflexible  polymer.

\acknowledgments
We thank S.\ Klumpp for discussions on the simple symmetric exclusion
process.


\end{document}